# Large tuneable exchange fields due to purely paramagnetically limited domain wall superconductivity

Pushpak Banerjee, Pramod K. Sharma, Sonam Bhakat, Biswajit Dutta & Avradeep Pal*

*Department of Metallurgical Engineering and Materials Science, Indian Institute of Technology, Bombay, Powai, Maharashtra – 400076.*

**Abstract**
The ability to locally apply and tune large magnetic fields is a crucial requirement for several devices, most notably for detection and generation of majorana fermions. Such a functionality can be achieved in Superconductor (S) /Ferromagnet (F) bilayers, where superconductivity is strengthened on top of domain walls due to local lowering of the proximity induced effective exchange fields. This is predicted to result in significant superconducting Tc enhancements and possible complete magnetic controlled switching on and off of the superconducting state. By using thin films of superconducting Nb and ferromagnetic insulating (GdN) bilayers, and through detailed magneto-transport measurements, we demonstrate the previously unobserved phenomena of complete switching in and out of the S state in S/F bilayers. In the thinnest of Nb layers, we estimate that the domain wall state induced tunability of proximity induced exchange fields can be as high as 1.3T with application of in plane external fields of only a few mT.

## 1. Introduction

Thin film S/F systems have been studied extensively due to the possibility of detection of novel interfacial effects and prospective dissipation less device functionalities[1]. Notable among these are the theoretically predicted and experimentally verified 0-pi transitions[2–4], resulting in an oscillatory order parameter; and the generation of long range odd frequency triplet correlations[5–8]. Despite the tremendous experimental advances over the last two decades, one of the longstanding theoretical predictions that has eluded experimental observation is that of a purely magnetic domain state mediated and purely paramagnetically limited switching of the superconducting state to the normal state and vice versa in S/F bilayers[9,10].

In S/F bilayers, Cooper pairs interact with F either through a proximity induced exchange field, causing paramagnetic pair breaking[9,10] or with stray magnetic fields emanating from F, resulting in orbital pair breaking[11]. As opposed to orbital pair breaking, which causes a heterogeneous superconducting region mixed with normal vortex cores[12,13]; paramagnetic pair breaking is especially useful for the tuning of induced exchange fields, with the superconductor remaining homogeneous, but with a reduced energy gap. When the underlying ferromagnet transitions into a multi-domain state, it results in areas over domain walls with reduced pair breaking exchange fields, thus resulting in a reduced average effective exchange field in the S layer. The difference in average exchange fields in the saturated and multidomain state of the ferromagnet is therefore tantamount to exchange field tunability in the S layer. Under appropriate conditions, this field tunability can result in a superconducting phase in the multi-domain state and a normal state (N) in the saturated state of the ferromagnet. This effectively translates into a measurable difference of Tc in the saturated and coercive fields states of the ferromagnet. In case a complete N-S transition is achieved, where the difference in Tc between the saturated and coercive states is greater than the broadness of the superconducting transition; the results can be compared with theoretically predicted N-S phase diagrams of high field superconductivity and paramagnetically limited domain wall superconductivity to estimate the tunability of exchange field in the system. Thus far, experiments investigating paramagnetic pair breaking in S/F bilayers has shown to cause incomplete and minimal N-S transitions[14–16], with very low differences in Tc (30mK or less), and hence magnitudes of exchange field tunability have not been derived. In this article, we fill up this experimental gap using Nb/GdN bilayers, by demonstrating a purely proximity induced exchange field modification driven N-S transition of thin film Nb, by applying small fields (few mT) which alters the domain state of a thin film soft ferromagnetic Insulator (FI) – GdN.

As opposed to a metallic ferromagnet, in S/FI systems, the situation is slightly different. When measured in Current-In Plane (CIP) geometry, with thin S layers (thickness of S layer lower than BCS bulk coherence length) electrons of the Cooper pair undergo reflections at the S/FI interface[17]. This again leads to an exchange interaction during the scattering event which manifests as an effective exchange field in the thin S layer. However, unlike S/F layers with metallic ferromagnets, a FI layer prevents diffusion of cooper pairs into itself, and therefore this constitutes a purely interfacial proximity induced exchange field effect[17–19]. The superconducting coherence length ($\xi_T$) close to transition abides by the Ginzburg Landau (GL) relation $\xi_T = \frac{\xi_{T=0}}{\sqrt{\frac{\Delta T}{T_C}}}$.

When the cooper pair interacts with FI moments in a single domain, all aligned by the Weiss-field, it experiences a uniform exchange field that contributes to the paramagnetic pair breaking effect and leads to a quasiparticle spin split superconducting state[20,21]. The cooper pairs can also sample the moments of two adjacent domains if $\xi_T$ lies in a range between domain wall width ($d_w$) and the modal lowest dimension of the domains. In this scenario, the effective exchange field would reduce, resulting in reduced paramagnetic pair breaking, which, as described before, is measurable through an enhanced $T_c$ of the system[9,10].

This phenomena of Domain Wall Superconductivity (DWS) although in spirit is very similar, but microscopically distinct from the spin switch effect in F/S/F trilayers[22], where switching the direction of the magnetic moment of the soft F layer in comparison to a harder F layer causes a $T_c$ enhancement due to an overall relative parallel and anti-parallel arrangements of the two ferromagnets which are approximated to be macro-spins, sans their finer domain texture. Moreover, as opposed to FI/S bilayers, where the field tuneable S layer can be interfaced with other device components (for example – a nanowire in case of Majorana devices); F/S/F trilayers may not useful for applications seeking local field tunability.

In addition to filling up the experimental gap of precisely measuring the exchange field tunability, and showing complete N-S transitions in an S/FI system; we also test most of the predictions of theory of paramagnetically limited DWS by controllably tuning the following intrinsic physical parameters: $\xi_T$, $\xi_{T=0}$, and domain wall area fraction. This can be done by varying the corresponding experimental parameters: temperature (T) for changing $\xi_T$, superconducting Nb film thickness[23], which affects $\xi_{T=0}$ in the dirty limit, and externally applying a magnetic field (H) which affects the domain state of the ferromagnet.

## 2. Results and Discussion

In Figure 1(a), we measure Resistance (R) as a function of increasing temperature at corresponding field values of externally applied magnetic field, from positive to negative saturation and vice versa. The sharp horn-like features are suggestive of how the $T_c$ enhancement is correlated to the coercive field of the FI. The $T_c$ enhancement can also be visualised from isolated R-T plots (vertical linecuts extracted from the colormap) for the disparate external magnetic field conditions namely Saturation field (-10mT), zero-field and coercive field conditions as shown in Figure 1(b). The RH plot in Figure 1(c) represents a horizontal linecut of the colormap at 4.2K, and shows a complete N-S transition, and therefore constitutes a demonstration of Infinite Magneto-Resistance (IMR). This IMR state coincides with the magnetization switching field values obtained from magnetic moment (M) vs H SQUID VSM measurements of a 3nm GdN film without a top Nb layer. Figure 1(d) shows an identical measurement performed on a sample where the interface between GdN and Nb is broken by a thin layer (3nm) of insulating AlN in between. This sample is grown in the same sputtering growth run, with the same plasma for GdN and Nb, and their thickness are identical to the one measured in Figure 1(a). However, we find that the $T_c$ of the system is far greater than without AlN in between, and the horn like features emblematic of GdN switching field dependent Tc enhancement are missing. This control experiment demonstrates the purely interfacial nature of the effect, and the minimal role of orbital pair breaking caused by stray fields from the FI layer.

While recording RT heating measurements, we start in the superconducting phase with relatively shorter cooper pairs, as compared to close to $T_c$. From saturation to the remnant field, the $T_c$ remains largely constant. This can be attributed to a mono-domain approximation (owing to observed squarish nature of MH in Figure 1(c)) causing an external field independent paramagnetic pair breaking effect on the Cooper pairs in this field region. While progressing towards the coercive state of the FI, a multi-domain state starts to set in, and interaction of Cooper pairs with adjacent domains becomes more probable with relatively larger $\xi_T$. Due to the opposite spin orientations in adjacent domains, and net zero spin in the domain walls, the effective exchange field in Cooper pairs is reduced. As the sample is heated in fields close to the switching field, we approach a temperature, where a sweet spot $\xi_T \cong d_w$ can be reached. This provides added stability to the cooper pairs by increasing the occurrence of adjacent domain sampling events (star marked events from Figure 1(e)) compared to intra-domain interaction events. At the MH switching fields, we have the largest areal fraction of domain walls, and lowest average domain size, and hence a maximum suppression of the effective exchange field leading to minimum pair breaking effects, thereby resulting in large $T_c$ enhancements (horn like features in Figure 1(a)). Further heating results in the condition $\xi_T \gg d_w$, and hence the effect diminishes as the cooper pairs become long enough to encounter pairs of distant domains, which may not conceal opposing moment orientations, and domain wall effects become negligible in this new spatial length scale. Thus, the effective exchange field experienced by the superconductor is less likely to be any lower. Beyond the switching fields, the domains again gain size, and the areal fraction of domain wall reduces drastically, thereby resulting in gradually lower $T_c$ as we approach the negative saturation field. Figure 1(e) is a cartoon representation of the observed effect, assuming in plane Neel type domain walls are prevalent in thin film GdN.

In Figure 2, we plot the dependence of various properties of these bilayers as a function of Nb thickness. Above 20 nm, no discernible horn like features were observed in colormaps, and below 7nm, the bilayers showed no superconducting transitions till the lowest temperatures at our disposal - 0.25K. Please refer to supplementary Figure 1 for colormaps of samples at these two extremes. Two noticeable aspects of thickness dependence present in Figure 2(a) are - the rapid decrease of the $T_c$ enhancement effect with increasing thickness (evident through both $\frac{\Delta T}{T_c} = \frac{T_{coer} - T_{sat}}{T_{sat}}$ and $\Delta T = T_{coer} - T_{sat}$ data, where $T_{coer}$, $T_{sat}$ are the transition temperatures at coercive field and saturation fields respectively) , and the reduced $T_c$ of Nb/GdN films compared to identical films layer thickness but with a thin 3nm of insulating AlN in between GdN and Nb (inset to Figure 2(a)). With reference to the later observation, we also note that the magnitude of $T_c$ decrease is highest for the thinnest Nb films. We attribute this to lowering of average proximity induced exchange fields with increasing Nb thickness[17]. We later go on to estimate the exchange field in our samples in Figure 3. The former observation is due to the reduction of the mean free path of Nb (l) which in turn reduces the dirty limit coherence length ($\xi_{T=0} \sim 0.855\sqrt{l\xi_0}$) of Nb with reduction in its thickness (cross symbol Figure 2(b)). Mean free path was estimated from both residual resistivity ($\rho_{10K}$) [23,24] as well as perpendicular to the plane critical field measurements (using relations $H_{C2\perp} = \frac{\phi_0}{2\pi\xi_T^2}$ and $H_{C2\perp} = H_{C2\perp}(0) \frac{1-(T/T_c)^2}{1+(T/T_c)^2}$ of Nb/GdN bilayers.

It is expected that the largest $T_c$ enhancements should occur when the GL coherence length close to $T_{coer}$ is comparable to domain wall width $d_w = \pi\sqrt{\frac{A}{K}}$ (where A is exchange stiffness coefficient and K the magnetocrystalline anisotropy of the FI layer). We estimate $d_w$ from several calculations of A and K available in the literature[25], and present it as a spread of values in Figure 2(b). We observe that the limit $\xi_T \cong d_w$ is indeed satisfied for the lowest thickness samples, and gradually settles to $\xi_T \gg d_w$. This observation therefore tallies with decreasing $\Delta T$ with increasing Nb thickness, as the effect of domain wall induced stability of Cooper pair is greatly reduced for larger sized Cooper pairs in thicker Nb samples.

We next turn our attention to the influence of the texture of the proximity induced exchange field in terms of its effect on the nature of S-N phase transition. This is best portrayed through the theory of localised non-uniform Domain Wall Superconductivity (DWS) put forth by Houzet-Buzdin (H-B)[10], as compared to the conventional uniform high field superconductivity treatments by Maki-Fulde (M-F)[26–28]. For this analysis, we trace out the phase diagram for domain wall induced non-uniform superconductivity[10], and superimpose our $T/T_{C0} = T_{Nb/GdN} / T_{\frac{Nb}{\frac{AlN}{GdN}}}$ data on it in Figure 3. We assume that $T = T_{sat}$ (blue star) corresponds to the MF curve, whereas $T = T_{coer}$ corresponds to the H-B curve and thereby extract the amplitude of proximity induced exchange field ($h_C$) from the MF curve and the hypothetical critical exchange field ($h_{CW}$) necessary for eliminating the domain wall superconducting phase[10] from H-B boundary. We find that although $h_C$ for all samples (blue stars in Figure 3) is of the order of several Tesla, it is always less than the internal molecular field for GdN ($T_{Curie} \approx 35K$) which we estimate as 17.3T from Curie-Weiss theory. As per quantities defined in H-B theory, in the DWS phase the area averaged exchange field ($h_{av} = f h_C$) where we define $f = \frac{h_C}{h_{CW}} < 1$, implying that the average exchange field has reduced from $h_C$, permitting a higher transition temperature at the switching fields ($T_{coer}$), however the amplitude remains at $h_C$. We find that this variation $(1-f)h_C$ of the effective proximity induced exchange field due to non-uniform magnetic texture can be as high as 1.3T (Nb(7)/GdN(3)) and reduces to 30mT (Nb(20)/GdN(3)) with increasing Nb film thickness.

We further investigate the in plane critical field behaviour of both Nb (7 and 8 nm)/GdN and Nb (7nm)/AlN/GdN films deep inside their respective superconducting states, and track the transition boundary (coloured circles in Figure 3). For Nb/GdN samples we use $\frac{h}{T_{C0}} = \frac{\mu_e\mu_0 H_{app}}{K_B T_{C0}} + \frac{h_c}{T_{C0}}$ (where $H_{app}$ is the magnitude of externally applied magnetic field in Tesla) and from our observations from Figure 1(e) we assume $h_C = 0$ for Nb/AlN/GdN. A second order broad transition is evident (refer to supplementary figure 2) at all temperatures for all samples and yet we observe a deviation from M-F second order phase boundary (blue dotted line in Figure 3). Such deviations are expected for superconductors with appreciable spin-orbit scattering (which includes Nb[29]) as predicted by Werthamer Helfand Hohenberg (WHH) theory[30]. Moreover, as per WHH, spin paramagnetic scattering and mean free path are inversely related and this translates to a higher critical field phase boundary for higher mean free path. We observe this exact phenomenon on contrasting the data of Nb(7)/GdN(3) (violet circles in Figure 3) against Nb(8)/GdN(3) (Indigo circles in Figure 3). Furthermore, in accordance to WHH predictions we breach the Chandrashekar-Clogston (CC) limit[31,32] for the Nb(8)/GdN (3) sample at low temperatures.

## 3. Conclusion

In conclusion, by using the novel approach of controlled Cooper pair shrinkage in the context of DWS - our results for the first time in S/F bilayers, demonstrate complete switching on and off of the superconducting state by purely paramagnetic effects of underlying FI magnetic microstructure. The above switching effect is tantamount to large tunable exchange fields (>1T) which are greatly significant for Majorana Zero Mode (MZM) devices where local field tunability with minimal external fields are necessary[33–35]. Our results demonstrate that small area ultrathin Nb/GdN bilayers are useful candidates for the purpose. Moreover, as per our knowledge, this study is the only attempt thus far to compare experimental data with phase diagram of paramagnetically limited DWS. Finally, such paramagnetically limited switchable bilayers can be used for designing cryogenic memory devices with switching fields which are approximately two orders of magnitude lower than that reported for orbitally limited DWS.

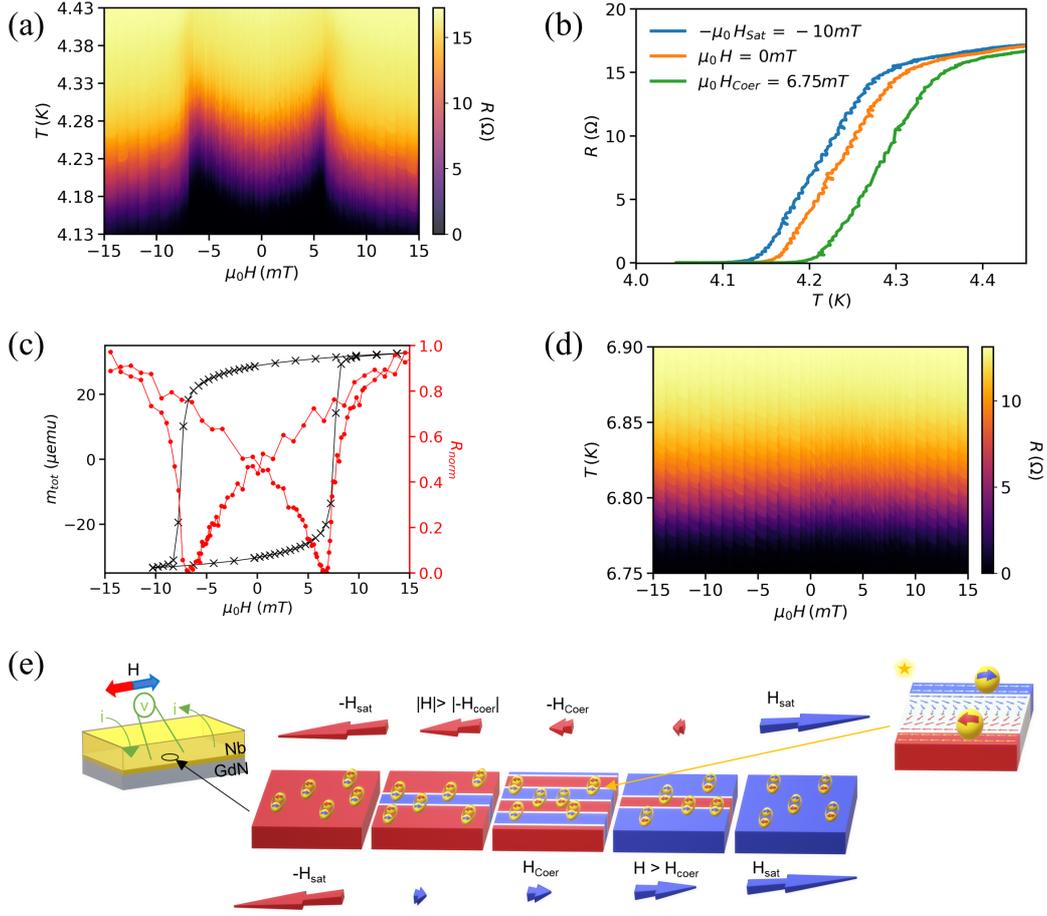

**Figure 1.** Observation of Tc enhancement of Nb and Infinite-Magnetoresistance in a Nb(8nm)/GdN(3nm) stack. **(a)** The measured Resistance R(Ω) of Nb as a function of applied external field μ₀H (mT) and Temperature T (K) represented as a composite of the colormaps for each half ($H_{sat} \rightarrow -H_{sat}$ and $-H_{sat} \rightarrow H_{sat}$) of the looped field sequence (Note: $H_{sat} \rightarrow 0$ and $-H_{sat} \rightarrow 0$ have been overlaid by $0 \rightarrow H_{Coer} \rightarrow H_{sat}$ and $0 \rightarrow -H_{Coer} \rightarrow -H_{sat}$ for visual clarity). The apparent horn like features at $\pm H_{coer}$ are characteristic of an enhanced $T_C$. **(b)** The RT plots for the three distinct field conditions ($H_{sat}$, zero field and $H_{coer}$) are displayed. **(c)** Normalized resistance '$R_{norm}$' as a function of applied external field μ₀H (mT) at 4.2K (red circles) and Magnetic moment $m_{tot}$ (μemu) vs. μ₀H (mT) (black cross) for a GdN (3nm) film at 4.2K (red and black connecting lines are a guide to the eye). **(d)** RT vs H colour plot with a field sequence of -15mT to 15mT for a Nb(8)/AlN(3nm)/GdN(3nm) lacking horn like $T_C$ enhancement features. **(e)** a cartoon illustration of DWS phenomena depicting the interaction of the cooper pairs with supposed micromagnetic structure at the interface of Nb (S) and GdN (FI). The spin projection of the Blue/Red colour stamped 'electron' of each pair and the magnetic moment orientation of the colour coded FI domains are matched with the 'H'. The star marked event showcases a particular event when $\xi_T \cong d_w$ and the net spin structure of the Neel domain wall (grey color) results in reduced exchange field on the cooper pair, strengthening the superconducting phase and leading to $T_C$ enhancement.

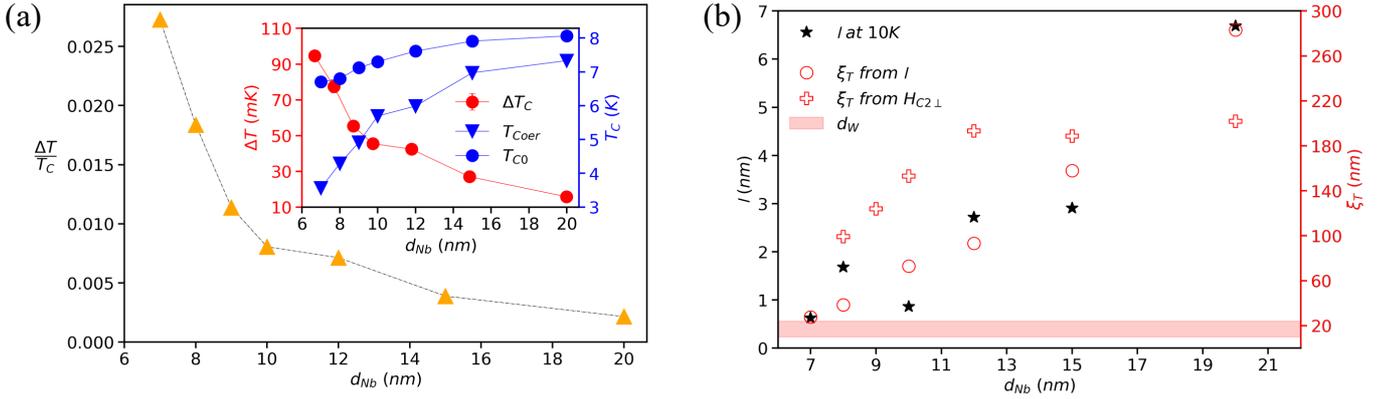

**Figure 2.** Relation between superconducting film thickness and enhanced transition temperature at coercive fields of FI. **(a)** $T_C$ enhancement factor $\frac{\Delta T}{T_C}$ (orange triangles) as a function of increasing Nb film thickness ($d_{Nb}$). Red circles in inset shows dependence of $\Delta T$ on $d_{Nb}$. Blue circles in inset represent the transition temperature ($T_{C0}$) of Nb/AlN/GdN while blue triangles represent switching field $T_C$ ($T_{Coer}$) of Nb/GdN stacks for various Nb thickness. All connecting lines are a guide to the eye. **(b)** Black stars represent the calculated mean free path (l) at 10K from residual resistivity measurements. Red circles are calculations of dirty limit GL coherence length ($\xi_T$) at coercive fields derived from $l$ and $\frac{\Delta T}{T_C}$. Red cross symbols are $\xi_T$ values derived from $H_{C2\perp}$ measurements.

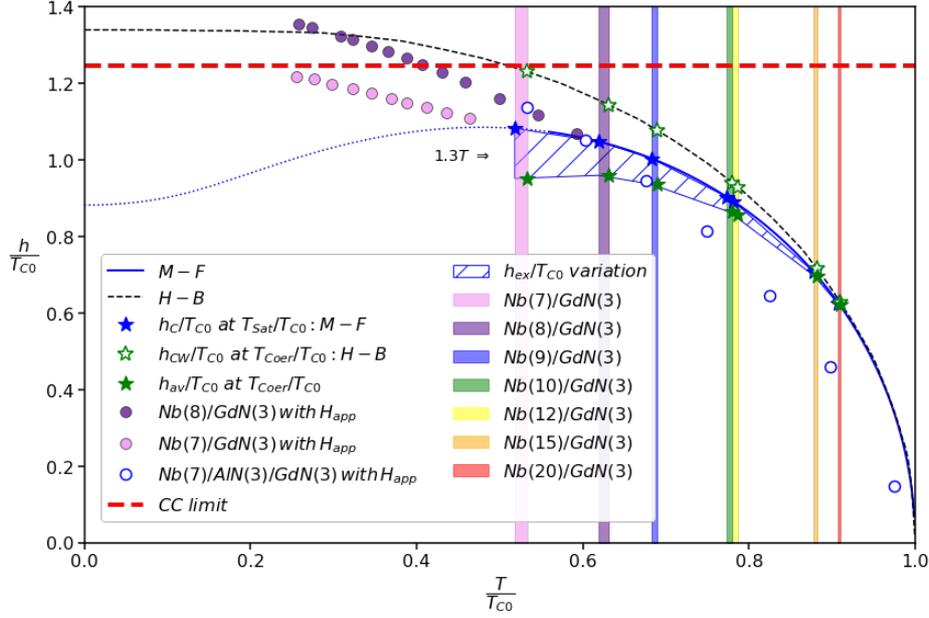

**Figure 3.** Estimating the effective exchange field encountered by the Superconducting layer in the S/FI system. The black dashed curve represents the H-B boundary for DWS (applicable to S/FI when FI is in coercive state at $\frac{T_{Coer}}{T_{C0}}$). The blue curve (FI in Saturated state at $\frac{T_{Sat}}{T_{C0}}$) is the M-F predicted phase boundary for uniform superconductivity. Coloured bands (VIBGYOR starting from left) represent the quantity $\frac{T_{Coer}-T_{Sat}}{T_{C0}}$ for Nb/GdN films having different Nb thickness (7-20nm). Blue filled stars correspond to $\frac{T_{Sat}}{T_{C0}}$ of Nb/GdN samples and therefore serve as estimates of critical exchange field ($\frac{h_C}{T_{C0}}$) on MF curve for uniform superconductivity. Green hollow stars represent $\frac{T_{Coer}}{T_{C0}}$ and are estimates for the required critical exchange field ($\frac{h_{Coer}}{T_{C0}}$) on the H-B for non-uniform superconductivity. Green filled stars are the derived average exchange fields ($\frac{h_{av}}{T_{C0}}$) at $\frac{T_{Coer}}{T_{C0}}$. The variation in exchange field ($\frac{h_{ex}}{T_{C0}}$ hatch shaded) due to DWS reduces (from about 1.3T for Nb(7))/GdN(3)) with increasing thickness. The data for Nb(7)/GdN(3) (violet circle) and the Nb(8)/GdN(3) (indigo circle) was acquired at high fields ($\frac{\mu_e\mu_0 H_{app}}{K_B T_{C0}}$) and low temperatures compared to their respective T$_{Sat}$ and have been offset along the y-axis by their respective $\frac{h_C}{T_{Sat}}$. The blue hollow circles trace the critical field vs temperature for a Nb(7)/AlN(3)/GdN(3) system and the highest point is of the order of 11.3T. The normalised paramagnetic limit (CC limit around 1.25T$_{C0}$) is marked in dashed red.

**Experimental Methods**

The S(Nb)/FI(GdN) were grown on n-doped Si substrates with a 285nm thick SiO$_2$ thermally grown oxide. AlN was used as a buffer layer for growth of GdN and as a capping layer, resulting in an overall multi-layered stack of the form - AlN(10nm)/ Nb(varying thickness)/ GdN (3nm)/ AlN(15nm). All layers were deposited via magnetron sputtering inside a custom-made multi-target Ultra High Vacuum chamber with a base pressure of the order 10$^{-9}$ mbar. The GdN layer was deposited by means of reactive dc sputtering of Gd in an 8% N$_2$ and 92% Argon atmosphere, while AlN films were obtained by reactive sputtering of Al in a 44% N$_2$ and 56% Ar mixture at 1.5 Pa. Nb thickness was varied in a single run by keeping samples in a rotating table which was rotated under the Nb targets using a computer controlled stepper motor. Magnetisation measurements as a function of in plane magnetic fields were performed on a AlN/GdN(3nm)/AlN sample using a M/s Quantum Design SQUID magnetometer system. For the magneto-transport measurements, the multi-layered stacks were wire bonded in a 4-point geometry, to a custom-made PCB devoid of any magnetic material, which were then affixed to a probe that was lowered into a variable temperature insert of an Oxford Teslatron Pulse tube Cryostat which could cool the sample stage down to 1.5K, and having the capability to apply uniaxial magnetic fields upto 12T. In plane magnetic fields were applied using the inbuilt superconducting solenoid of the Teslatron system. The QCODES platform was adopted for data acquisition.

**Acknowledgements**

We thank Hridis K. Pal, Department of Physics, IIT Bombay, for valuable discussions. We express our sincere thanks to Dr. R. J. Choudhary, UGC-DAE CSR, Indore for helping us carry out the SQUID VSM measurements. The work was financially supported by a Core Research Grant from Department of Science and Technology, India, file number CRG/2019/004758.

**Author contributions**

PKS, PB and AP conceived the experiment. PB grew the multi-layered samples, performed the low temperature magneto-transport measurements. SB optimised the thin film growth recipes for Nb and GdN. BD performed SQUID VSM measurements. PKS fabricated hall bars on the multilayers for residual resistivity measurements. PB and AP analysed the data and wrote the manuscript. All authors discussed the results and commented on the manuscript.